\documentclass[preprint,showpacs,amssymb,aps,prd]{revtex4}% Preprint
\begin{document}

\def\diagram#1{{\normallineskip=8pt
       \normalbaselineskip=0pt \matrix{#1}}}

\def\diagramrightarrow#1#2{\smash{\mathop{\hbox to
.8in{\rightarrowfill}}
        \limits^{\scriptstyle #1}_{\scriptstyle #2}}}

\def\diagramleftarrow#1#2{\smash{\mathop{\hbox to .8in{\leftarrowfill}}
        \limits^{\scriptstyle #1}_{\scriptstyle #2}}}

\def\diagramdownarrow#1#2{\llap{$\scriptstyle #1$}\left\downarrow
    \vcenter to .6in{}\right.\rlap{$\scriptstyle #2$}}

\def\diagramuparrow#1#2{\llap{$\scriptstyle #1$}\left\uparrow
    \vcenter to .6in{}\right.\rlap{$\scriptstyle #2$}}

%%%%%%%%%%%%%%%%%%%%%%%%%%%%%%%%%%%%%%%%%%%%%%%%%%%%%%%%%%%%%%%%%%%%%%%%%%%%%
%
%%%%%%%%%%%%%%%%%%%%%%%%%%%%%%%%%%%%%%%%%%%%%%%%%%%%%%%%%%%%%%%%%%%%%%%%%%%%%
\title{Anomalous Commutator
Algebra
\\
for Conformal Quantum Mechanics}
\author{Gino N. J. A\~{n}a\~{n}os,$^{1,2}$
 Horacio E. Camblong,$^{3}$ \\
   Carlos Gorrich\'ategui,$^{1}$  Ernesto Hern\'andez,$^{1}$
and
Carlos R. Ord\'{o}\~{n}ez$^{1,2}$}

\affiliation{
$^1$ Department of Physics, University of Houston, Houston,
TX 77204-5506
\\
$^2$
World Laboratory Center for Pan-American Collaboration in Science and
Technology,
\\
University of Houston Center, Houston, TX 77204-5506
\\
$^3$
Department of Physics, University of San Francisco, San
Francisco, CA 94117-1080}

\begin{abstract}
The structure of the commutator algebra for conformal quantum mechanics is
considered.  Specifically, it is shown that
the emergence of a dimensional scale by renormalization implies
the existence of an anomaly or quantum-mechanical symmetry
breaking, which is explicitly displayed at the level of the generators of the
SO(2,1) conformal group.
Correspondingly,
the associated breakdown of the conservation of the dilation and special
conformal charges is derived.
\end{abstract}

\maketitle

\section{Introduction}
\label{sec:introduction}

The concept of symmetry plays a central role in the conceptual framework 
of modern physics. One of the most fruitful
approaches starts by identifying a symmetry, whose actual or potential breakdown is 
subsequently analyzed. A particular
case of this process is an anomaly---a classical symmetry that breaks down upon regularization and
 renormalization~\cite{a-b-j,tre:85,jac:update_anomalies}.
Specifically, the existence of anomalies is usually associated 
with the need to regularize
 infinities that appear in quantum-field-theory descriptions of particle and
 extended-object interactions. 
In their continuum version, these theories
 require an infinite number of degrees of freedom,
 which in turn become the source of
 infinities in relevant calculations.
As a consequence, 
regularization is unavoidable and occasionally accompanied by 
anomalous symmetry breaking. 

In contrast with their quantum-field-theory counterparts, the concepts of 
regularization, renormalization, and anomalous symmetry breaking
  do not appear to be necessary tools
in quantum mechanics.
This ``regular'' behavior is usually ascribed to 
the finite number of degrees of freedom 
sufficient to describe these systems at low energies.
 However, this lore has been
 challenged by Jackiw~\cite{jac:91} 
for the two-dimensional
 $\delta$-function interaction.
The strongly singular nature of
 this potential at the origin suggests the use of regularization
 and renormalization as an alternative approach to quantizing the system, which
 would otherwise seem not to be defined.
In fact, a detailed calculation for
the two-dimensional
 $\delta$-function potential
shows that the interacting system is well defined, 
but {\em only\/} after renormalization~\cite{jac:91,tho:79}.
The existence of a renormalized version of the
theory and the usefulness of related field-theory concepts have been confirmed
 in a number of independent studies~\cite{delta_biblio,cam:dtI, cam:dtII}.
Moreover, 
a simple argument reveals that
this interaction 
is scale invariant~\cite{cam:dtI}, but
 a dimensional parameter
 survives regularization and 
renormalization
due to
dimensional
transmutation~\cite{col:73}.
In addition,
similar techniques and concepts have been used to analyze and renormalize
the inverse square potential~\cite{gup:93,isp_letter,cam:dtI,cam:dtII,beane:00},
which can also be shown 
to be conformally invariant at the classical level~\cite{jac:72,alf:76}.
These scale-invariant potentials
may be regarded as the most outstanding examples of conformal quantum mechanics.

 Despite our recent progress in the analysis of singular potentials
in conformal quantum mechanics,
the possible breakdown of their symmetry algebra
at the quantum level has not yet been systematically explored.
This omission is corrected in the present  paper, 
in which we introduce an outline of the general theory
and detailed computations for the particular case of 
 the two-dimensional  $\delta$-function interaction.
The general theory is presented in
Sec.~\ref{anomalous_comm_alg},  where 
we review the algebraic commutator properties 
of the Noether charges associated with 
conformal invariance
and show how to characterize the corresponding
anomaly at the quantum level.
In Sec.~\ref{sec:2D_delta} we
display the emergence of the conformal anomaly for
 the two-dimensional $\delta$-function potential  by
implementing the necessary renormalization of the bound-state sector
using three
different regularization techniques.
In Sec.~\ref{sec:scattering}
we also show how to describe this 
anomalous symmetry breaking for the scattering sector of the theory.
Finally, in Sec.~\ref{sec:conclusions}
we present the conclusions of our work,
while in the Appendix
we summarize the main results on the $d$-dimensional radial
 Schr\"odinger equation that are needed throughout the paper.

%%%%%%%%%%%%%%%%%%%%%%%%%%%%%%%%%%%%%%%%%%%%%%%%%%%%%%%%%%%%%%%%%%%%%%%%%%%%%
%
%%%%%%%%%%%%%%%%%%%%%%%%%%%%%%%%%%%%%%%%%%%%%%%%%%%%%%%%%%%%%%%%%%%%%%%%%%%%%
\section{Anomalous Commutator Algebra}
\label{anomalous_comm_alg}

In this section
we consider an arbitrary scale-invariant potential
$V({\bf r})$.  
From
a simple dimensional argument~\cite{cam:dtI}, 
it can be shown that 
 the scale invariance of the action occurs if and only if 
$V({\bf r})$ is
homogeneous of degree $-2$.
In subsequent sections of this paper,
$V({\bf r})$ will be specialized
to the particular case of the 
two-dimensional 
$\delta$-function interaction.

\subsection{SO(2,1)  commutator algebra}
\label{SO(2,1)_comm_alg}

 A straightforward analysis of the symmetries of these scale-invariant
potentials
under time reparametrizations
shows the existence 
of  three Noether charges.
The corresponding quantum-mechanical generators are
the Hamiltonian
\begin{equation}
N_{1}=H
\equiv
 \frac{p^{2}}{2M}+V({\bf r})
\; ,
\label{eq:Hamiltonian}
\end{equation}
the dilation operator
\begin{equation}
N_{2}=D
\equiv
tH
- \frac{1}{4}
\left( {\bf p} \cdot {\bf r}
+  {\bf r} \cdot {\bf p}
\right)
\;  ,
\label{eq:dilation_op}
\end{equation}
and the special conformal operator
\begin{equation}
N_{3}=K
\equiv
t^{2}H
-
\frac{t}{2}
\left( {\bf p} \cdot {\bf r}
+  {\bf r} \cdot {\bf p}
\right)
+ \frac{M}{2}
r^{2}
\;  ,
\label{eq:special_conformal_op}
\end{equation}
which
are expected to satisfy the
SO(2,1) Lie algebra~\cite{wyb:74}
\begin{equation}
[D,H]_{\rm regular}
= - i \hbar H
 \;  ,
\;
\;  \;   \;   \;
[K,H]_{\rm regular}  = - 2 i \hbar D
\;  ,
\;
\;  \;   \;   \;
[D, K]_{\rm regular}  =  i \hbar K
\;  .
\label{eq:naive_commutators}
\end{equation}
In Eq.~(\ref{eq:naive_commutators})
the qualification ``regular'' emphasizes that
the commutators follow from a naive computation in which their
anomalous behavior is not explicitly considered.
These scaling and commutator
properties 
have been shown to apply to the 
 two-dimensional
 $\delta$-function interaction~\cite{jac:91} 
and the
inverse square potential~\cite{jac:72,alf:76},
and are also shared 
by
the magnetic monopole~\cite{jac:80}
and the magnetic vortex~\cite{jac:90}.
These representative examples
of conformal quantum mechanics and their associated SO(2,1) symmetry
have also
been recognized in the study of a number of nonrelativistic limits
of quantum field theories~\cite{jac:90b,bergman}.
More generally,
the same basic results apply to the entire class of
homogeneous potentials of degree $-2$, which are 
both scale and conformally invariant.

The usual interpretation of scale invariance is
summarized by the first commutator in Eq.~(\ref{eq:naive_commutators}),
which shows that the scale dimension of $H$ is $-1$.
This scale dimension is, in fact, the ``time dimension'' ${\mathcal T}$
corresponding to the dimensional-analysis
result $[H]={\mathcal T}^{-1}$, in units such that $\hbar =1$ and $2M=1$.
 For an arbitrary interaction, in these units,
the spatial length dimension is
${\mathcal L}={\mathcal T}^{1/2}$.
Moreover, for the particular case of a scale-invariant
theory, i.e., for a potential $V({\bf r})$ homogeneous of degree
$-2$, the ``naive algebra'' and scaling
 $[H]={\mathcal T}^{-1}$ are
satisfied directly from the symmetry,
when other subtleties are ignored~\cite{cam:dtI}.
That is precisely the dimensional-analysis interpretation
of  the first commutator in Eq.~(\ref{eq:naive_commutators}).

The main goal of our  paper is to
 show the emergence of correction terms
in Eq.~(\ref{eq:naive_commutators})
due to
dimensional
transmutation~\cite{col:73},
as manifested
by the presence
of a dimensionful renormalization 
parameter~\cite{gup:93,isp_letter,cam:dtI,cam:dtII}.
In fact, this is the origin
of an experimental realization
  of a quantum anomaly in molecular physics~\cite{dipole_anomaly},
whose underlying mechanism
has also been studied within a path-integral 
approach~\cite{pi_singular,pi_delta,green}.

\subsection{Anomalous commutators}
\label{anomalous_comm}

In what follows we will consider the
nonperturbative definition of the Hilbert space for conformally-invariant 
potentials, according to the 
framework of Refs.~\cite{jac:91,cam:dtI,cam:dtII,isp_letter,pi_delta}.
This can be achieved by properly renormalizing
the theory in the strong-coupling regime, with the introduction
of a scale-breaking parameter.
However, if the SO(2,1)  conformal
symmetry is violated upon renormalization, then 
the question arises as to where this 
symmetry breaking manifests in the commutator algebra of the 
operators~(\ref{eq:Hamiltonian})--(\ref{eq:special_conformal_op}).
In this paper
 we show that the modification of the algebra~(\ref{eq:naive_commutators})
is encoded in the basic commutator
\begin{equation}
[D,H] =
- i \hbar H
+ [D,H]_{\rm extra}
\; ,
\label{eq:DH_commutator_anomalous}
\end{equation}
which acquires an ``extra'' piece 
whose expectation values will be computed below.
Then,
Eqs.~(\ref{eq:Hamiltonian})--(\ref{eq:special_conformal_op})
imply that
\begin{equation}
[K,H]
=
- 2 i \hbar D
+2t \,
[D,H]_{\rm extra}
\;  ,
\;
\;  \;   \;   \;
\left[D,K \right]
 =
 i \hbar K
- t^{2} \left[D,H \right]_{\rm extra}
\;  ,
\label{eq:KH_DK_commutator_anomalous}
\end{equation}
as follows 
by straightforward application
of the canonical commutator relations.
Thus,
the extra terms in
Eqs.~(\ref{eq:DH_commutator_anomalous}) and (\ref{eq:KH_DK_commutator_anomalous})
extend the commutator algebra~(\ref{eq:naive_commutators})
and spoil the conservation laws of the Noether 
charges~(\ref{eq:dilation_op}) and (\ref{eq:special_conformal_op}),
as dictated by their  
time evolution 
\begin{equation}
\frac{dA}{dt} 
=
\frac{ \partial A}{ \partial t} 
+
\frac{1}{i\hbar}
\left[
A, H
\right]
\;  
\label{eq:rate_of_change}
\end{equation}
in the Heisenberg picture.
In particular, the 
  Heisenberg equations~(\ref{eq:rate_of_change})
imply that 
\begin{equation}
\frac{dD}{dt}
=
\frac{1}{i \hbar}
\,
 [D,H]_{\rm extra}
\;  ,
\;
\;  \;   \;   \;
\frac{dK}{dt}
 =
\frac{2t}{i \hbar}
\,
[D,H]_{\rm extra}
\label{eq:D_K_conservation_law_failure}
\;  .
\end{equation}

The next step in our
construction is the remarkable finding that the
modified commutator algebra and the corresponding
relations~(\ref{eq:D_K_conservation_law_failure})
  can be evaluated in a representation-independent
manner. This is achieved again by the use of the canonical commutators, which imply
that
\begin{equation}
\left[
 {\bf p} \cdot {\bf r}
+  {\bf r} \cdot {\bf p}
, H
\right]
= 2 i \hbar
\left\{
2 T -
{\mathcal E}_{\bf r}   V ({\bf r})
\right\}
\; ,
\label{eq:Lambda_commutator}
\end{equation}
where $T=H-V$ is the kinetic energy operator
and the symbol
\begin{equation}
{\mathcal E}_{\bf r}
=
{\bf r} \cdot 
{\bf \nabla} 
\; 
\label{eq:Eulerian_derivative}
\end{equation}
stands for the ``Eulerian derivative,''
which---when applied to a
homogeneous function---selects
 the correct degree of homogeneity.
The expression ${\mathcal E}_{\bf r} V ({\bf r})$ 
is a formal
operator derivative that coincides with the corresponding 
elementary counterpart in the position representation.
Then, 
the ``anomaly operator''  
${\mathcal A} ({\bf r})$,
to be defined from the extra term  $[D,H]_{\rm extra}$
in Eq.~(\ref{eq:DH_commutator_anomalous}),
can be computed 
from Eqs.~(\ref{eq:dilation_op}) and (\ref{eq:Lambda_commutator}),
whence
\begin{equation}
{\mathcal A} ({\bf r})
\equiv
\frac{1}{i \hbar}
[D,H]_{\rm extra}
\equiv
\frac{1}{i \hbar}
[D,H] +  H
=
\left[
\openone
+
\frac{1}{2}
\,
{\mathcal E}_{\bf r}
 \right]
V ({\bf r})
\;  ,
\label{eq:time_rate_of_dilation_op}
\end{equation}
where $\openone$
is the identity operator.
An alternative useful expression of this 
anomaly~(\ref{eq:time_rate_of_dilation_op})
 in $d$ dimensions is
\begin{equation}
{\mathcal A} ({\bf r})=-\frac{1}{2} ( d - 2 )\, V ({\bf r})
+
\frac{1}{2}
\,
{\bf \nabla}
\!
\cdot
\!
\left\{
{\bf r} \,
 V ({\bf r})
\right\}
\;  .
\label{eq:time_rate_of_dilation_op_ddim}
\end{equation}

Despite their deceivingly simple appearance,
Eqs.~(\ref{eq:time_rate_of_dilation_op}) and
 (\ref{eq:time_rate_of_dilation_op_ddim}) 
still fail to make the anomalous behavior 
explicit.
This is due to the fact that the 
conformal anomaly
can be reduced to the breakdown of the 
naive scaling properties of the potential.
This fact is clearly displayed by 
Eq.~(\ref{eq:time_rate_of_dilation_op}), which
shows that the dilation charge is conserved and scale invariance is
maintained when
$
{\mathcal E}_{\bf r} V ({\bf r})
= -2 \, V ({\bf r})
$, 
an equation that amounts to
 Euler's theorem for a homogeneous potential of degree $-2$.
Thus, one is naively tempted
to state that the conformal anomaly vanishes
for the class of scale-invariant potentials.
However, as we will show below, this  homogeneity condition is violated:
{\it the breakdown of Euler's theorem can be traced to the singular behavior of
the potential and its associated wave functions at the origin}.
In particular, the existence of a nonvanishing anomaly 
can be explicitly shown
by considering the corresponding
expectation values with
normalized states $\left| \Psi \right\rangle$.

In subsequent sections we are going  to 
apply these generic concepts 
to the 
two-dimensional 
$\delta$-function interaction.
More precisely,
we will show that
the expressions in 
Eqs.~(\ref{eq:D_K_conservation_law_failure}) and
(\ref{eq:time_rate_of_dilation_op})--(\ref{eq:time_rate_of_dilation_op_ddim-EV})
are indeed nontrivial
due to the singular behavior of the wave function at the origin.
 This means that:
(i)
the additional term 
$[D,H]_{\rm extra}$
in the commutator
$[D,H]$, as defined in
Eq.~(\ref{eq:DH_commutator_anomalous}),
 is not identically equal to zero;
(ii)
relevant
expectation values of this extra term
$[D,H]_{\rm extra}$ have nonzero values.
Furthermore, 
this program can be most easily accomplished by computing
the expectation value
\begin{equation}\label{dddt}
\frac{d}{dt}
\left\langle
D
\right\rangle_{\scriptstyle \!  \Psi}
=
\left\langle
{\mathcal A} ({\bf r})
\right\rangle_{\scriptstyle \!  \Psi}
=
-
\frac{1}{2}
( d - 2 )
\left\langle
 V ({\bf r})
\right\rangle_{\scriptstyle \!  \Psi}
-
\frac{1}{2}
\int
d^{d} {\bf r}
\,
V ({\bf r})
\,
{\mathcal E}_{\bf r}
\left|
\Psi ({\bf r})
\right|^{2}
\;  ,
\label{eq:time_rate_of_dilation_op_ddim-EV}
\end{equation}
in which a vanishing boundary term at infinity is dropped,
after integration by parts.

\subsection{Properties of symmetry generators and their expectation values}
\label{symmetry_generators}

Before 
applying Eq.~(\ref{eq:time_rate_of_dilation_op_ddim-EV})
and related concepts to particular potentials, 
we will first summarize a number of 
well-known albeit insightful results about quantum-mechanical
expectation values.
These will help interpret the values taken by the conformal anomaly
within a familiar framework.

Specifically,
consider a 
 generator $A$ of a {\em symmetry\/} 
that satisfies the nontrivial
condition 
$
\partial A/\partial t \neq 0
$.
Let us also assume that
 a mechanism is provided for the existence of a state
$\left| \Psi \right\rangle$
that yields a nonvanishing expectation value
\begin{equation}
\left\langle
\frac{\partial A}{ \partial t}
\right\rangle_{\scriptstyle \!  \! \Psi}
\neq 0
\;  .
\label{eq:EV_time_derivative_of_A}
\end{equation}
For the important cases considered in this paper, 
 $A=D$ or $A=K$,
this mechanism happens to be renormalization. 
Then,
from general properties of quantum-mechanical states,
the following statements can be made:
\renewcommand{\theenumi}{(\arabic{enumi})}
\renewcommand{\labelenumi}{\theenumi}
\noindent
\begin{enumerate}
\item
If the
symmetry is strictly  maintained, then
$
-i \hbar
\,
 \partial A/\partial t
 =
[A,H]
 \neq 0
$, so that Eq.~(\ref{eq:EV_time_derivative_of_A}) implies 
that
$\left\langle
[A,H]
\right\rangle_{\scriptstyle \!  \Psi}
 \neq 0
$.
This is precisely what would happen with the dilation charge 
$D$ and conformal charge
$K$, if the commutators were exactly given by Eq.~(\ref{eq:naive_commutators}).

Reciprocally,
this statement is logically equivalent to the condition that, if
a state  $\left| \Psi \right\rangle$ is found for which
$\left\langle
[A,H]
\right\rangle_{\scriptstyle \!  \Psi}
= 0
$,  then the symmetry is necessarily broken.

\item
If there exist
{\em normalized stationary states\/}
 $\left| \Psi \right\rangle$,
then 
\begin{equation}
\left\langle 
[A,H]
\right\rangle_{\scriptstyle \!  \Psi}
= 0
\end{equation}
is also a necessary
condition. 
As a consequence,
 when
 Eq.~(\ref{eq:EV_time_derivative_of_A}) 
is satisfied,
 the symmetry is violated, with
\begin{equation}
\frac{d}{dt}
\left\langle
A
\right\rangle_{\scriptstyle \!  \Psi}
=
\left\langle
\frac{\partial A}{ \partial t}
\right\rangle_{\scriptstyle \!  \! \Psi}
\;  .
\label{eq:symmetry_violation_normalized_states}
\end{equation}

For the dilation and conformal charges,
this is only possible through the extra piece in
Eq.~(\ref{eq:DH_commutator_anomalous}),
which should guarantee a subtle chain of identities
\begin{equation}
i \hbar 
\,
\left\langle
\frac{\partial A}{ \partial t}
\right\rangle_{\scriptstyle \!  \! \Psi}
=
\left\langle [A,H]_{\rm extra}
\right\rangle_{\scriptstyle \!  \Psi}
=
-
\left\langle 
\; 
[A,H]_{\rm regular} 
\right\rangle_{\scriptstyle \!  \Psi}
\; ,
\end{equation}
where each individual term is not zero.
\end{enumerate}

In particular, the scheme discussed in point 2 above
applies directly to the ground state
 $
\left|
\Psi_{\! {\rm (gs)}}
\right\rangle
$, whenever it exists.
Then, Eq.~(\ref{eq:symmetry_violation_normalized_states})
[with $A=D$ defined in Eq.~(\ref{eq:dilation_op})]
implies that
\begin{equation}
\frac{d}{dt}
\left\langle 
D 
\right\rangle_{\scriptstyle \!  \Psi_{\rm \! {\scriptscriptstyle (gs)}}  }
=
E_{{\rm (gs)}}
\;  ,
\label{eq:time_rate_of_dilation_op-EV}
\end{equation}
where
$E_{{\rm (gs)}}
=
\left\langle 
H 
\right\rangle_{\scriptstyle \!  \Psi_{\rm \! {\scriptscriptstyle (gs)}}  }
$.

In the next few sections,
we will verify Eq.~(\ref{eq:time_rate_of_dilation_op-EV})
for the
two-dimensional 
$\delta$-function interaction
by an explicit computation of the anomalous correction
terms. 
In other words,
we will show that this potential
exhibits a conformal quantum anomaly.

\section{The Two-Dimensional 
$\mbox{\boldmath $\delta$}$-Function 
Interaction}
\label{sec:2D_delta}

In this section
we will consider a
 two-dimensional
$\delta$-function interaction
\begin{equation}
V({\bf r}) =  
g \, 
\delta^{(2)}  ({\bf r} ) 
\equiv 
-  
\frac{\hbar^{2}}{ 2M} 
\,
  \lambda
\,
\delta^{(2)}  ({\bf r} ) 
\; .
\label{eq:delta_interaction}
\end{equation}
In addition to defining 
the interaction potential,
Eq.~(\ref{eq:delta_interaction})
introduces a dimensionless coupling $\lambda$.
For the interaction~(\ref{eq:delta_interaction}),
the conformal anomaly defined in
Eq.~(\ref{eq:time_rate_of_dilation_op_ddim-EV}),
with $d=2$, is given by the formal expression
\begin{equation}
\frac{d}{dt}
\left\langle D 
\right\rangle_{\scriptstyle \!  \Psi}
=
- \frac{g}{2}
\int
d^{2} {\bf  r}
\,
\delta^{(2)} ({\bf r})
\,
{\mathcal E}_{\bf r}   
\left| 
\Psi ({\bf r})
\right|^{2}
\;  .
\label{eq:DELTA_time_rate_of_dilation_op_ddim-EV}
\end{equation}
 As we will see next,
Eq.~(\ref{eq:DELTA_time_rate_of_dilation_op_ddim-EV}) is ill defined 
and requires an appropriate
procedure of regularization and renormalization,
to be performed simultaneously with
the determination of states and observables.

The conformal anomaly 
discussed in this paper is manifested by the existence of a nonzero
value for the
right-hand side of
Eqs.~(\ref{eq:D_K_conservation_law_failure}),
(\ref{eq:time_rate_of_dilation_op})--(\ref{eq:time_rate_of_dilation_op_ddim-EV}),
and
(\ref{eq:DELTA_time_rate_of_dilation_op_ddim-EV}).
For the two-dimensional delta-function interaction,
a naive argument would suggest that this time derivative in
Eq.~(\ref{eq:DELTA_time_rate_of_dilation_op_ddim-EV})
is indeed identically equal to zero, because 
$\delta^{(2)} ({\bf r})$ selects a zero value in 
$ {\bf r} \cdot  {\bf \nabla} $. 
However, this line of reasoning assumes that the states  $\left| \Psi \right\rangle$
have a regular behavior 
at the origin---a condition that is explicitly violated upon renormalization
in the presence of the interaction~(\ref{eq:delta_interaction}).
More precisely,
even though the {\em regularized\/} wave functions satisfy the regular
boundary conditions~(\ref{eq:asympt_BC}) and (\ref{eq:BC_at_origin}),
the {\em renormalized\/}
 wave functions 
acquire a logarithmic singularity at the origin.
For example,
the Hilbert subspace of normalized bound-states
of the 
two-dimensional
$\delta$-function interaction
 reduces to
the one-dimensional space
spanned by the renormalized
ground state~\cite{jac:91,cam:dtI,cam:dtII,pi_delta}
\begin{equation}
\Psi_{\! {\rm (gs)}}
 ({\bf r})
=
\frac{\kappa}{ \sqrt{\pi} } \, K_{0} (\kappa r)
 \; ,
\label{eq:delta_wf_normalized_renormalized}
\end{equation}
where
\begin{equation}
E_{{\rm (gs)}}
=
-
\frac{\hbar^{2}\kappa^{2} }{2M}
\;  ,
\label{eq:GS_energy}
\end{equation}
while the running 
coupling constant $g$ asymptotically vanishes.
Specifically,
the behavior of the wave function~(\ref{eq:delta_wf_normalized_renormalized})
near the origin is dictated by~\cite{abr:72}
\begin{equation}
K_{0}(z)  
 \stackrel{(z \rightarrow 0)}{=}
 - \left[ \ln \left( \frac{z}{2} \right)
 + \gamma \right]
\,
\left[ 1 +O(z^{2})  \right] 
\;  ,
\label{eq:Mcdonald_0_small_arg}
\end{equation}
where  $\gamma$ stands for the Euler-Mascheroni constant. 
Thus,
 the integral in Eq.~(\ref{eq:DELTA_time_rate_of_dilation_op_ddim-EV})
fails to vanish identically and confirms the purported violation of  
Euler's theorem.
As a consequence,
Eq.~(\ref{eq:DELTA_time_rate_of_dilation_op_ddim-EV}) 
is ill defined at the level
of the renormalized quantities,  but can be evaluated by going back
to the regularized theory and taking the appropriate limit of its regularized
counterpart.

In what follows we will regularize 
Eq.~(\ref{eq:DELTA_time_rate_of_dilation_op_ddim-EV}) using
 three distinct techniques---and each one involves
defining the regularized potential, 
as well as the corresponding running
coupling constant $\lambda$:
(i) real-space regularization with a circular-well potential;
(ii)
 real-space regularization with a radial $\delta$-function potential;
and
(iii)
 dimensional regularization.
As we will see, subtle cancellations within each one of the 
regularization
methods combine to reproduce the same final answer~(\ref{eq:time_rate_of_dilation_op-EV}).

\subsection{Real-space
regularization with a circular-well potential}
\label{sec:circular_well}

Real-space regularization
provides a  scheme  
whereby the short-distance physics is appropriately modified 
for $ r \alt a$
(where
$a$ is a real-space regulator), so as to
yield a well-defined problem.
Of the many possible real-space regularization 
techniques,
here it proves convenient to introduce 
a  circular-well
potential
\begin{equation}
V({\bf r})  
\sim 
g
\,
\frac{  \theta (a-r) }{ \pi a^{2} }
\equiv
-
\frac{ \hbar^{2}\, \lambda}{2M}
\,
\frac{  \theta (a-r) }{ \pi a^{2} }
\;  ,
\label{eq:delta_interaction_circular_well}
\end{equation}
in which  $\theta (\xi) $ stands for the Heaviside function.

Due to the central nature of Eq.~(\ref{eq:delta_interaction_circular_well}),
the results summarized in the
Appendix
can be directly applied.
The corresponding
 Schr\"odinger equation 
for the reduced radial wave function $u_{l}(r)$ 
is given by
 \begin{equation}
\left\{
 \frac{d^2}{dr^2}
+ 
\left[ 
\frac{2M}{\hbar^{2}} \,   E +
\lambda  % (a)
\,
\frac{  \theta (a-r) }{ \pi a^{2} }
 \right]
-
\frac{l^{2} -1/4 }{r^2}
\right\}
u_{l} (r)
=
0
\;  ,
\label{eq:radial_Schr_2D_delta_circular_well}
\end{equation}
in which $l=|m|$, where $m$ is the usual 
quantum number.
The bound-state solution ($E<0$)
to Eq.~(\ref{eq:radial_Schr_2D_delta_circular_well})
can be written in terms of 
 Bessel functions~\cite{abr:72},
\begin{equation}
  R_{l}(r)
\equiv
\frac{ u_{l}(r) }{\sqrt{r}}
=\left\{\begin{array}{lr}
    \mbox{\boldmath\large  $\left\{  \right.$ } \! \! \!
J_{l} (\tilde k r)
\mbox{\boldmath\large  $,$ } \!   \!
N_{l} (\tilde k r)
\mbox{\boldmath\large  $\left.  \right\}$ } \! \!
\;  
& \textrm{for}
\;\; r<a  \; ,
\\
\mbox{\boldmath\large  $\left\{  \right.$ } \! \! \!
I_{l}(\kappa r)
\mbox{\boldmath\large  $,$ } \!   \!
 K_{l} (\kappa r) 
\mbox{\boldmath\large  $\left.  \right\}$ } \! \!
\;  
 & \textrm{for} \;\; r>a \; ,
\\
  \end{array}
\right.
\;   \;  
\label{eq:Bessel_solution_2D_delta_circular_well}
\end{equation}
where the curly brackets 
%the symbol
%$\mbox{\boldmath\large  $\left\{  \right.$ } \! \! \!
%\mbox{\boldmath\large  $,$ } \! \! \!
%\mbox{\boldmath\large  $\left.  \right\}$ } \! \!
%$
stand for linear combination, while
\begin{equation}
\tilde{k}^2
= 
\frac{2M}{\hbar^{2}} \,   
E
+\frac{\lambda}{\pi a^2} 
\;  
\label{eq:wave_number_2D_delta_circular_well}
\end{equation}
and
\begin{equation}
 \kappa^2
= - 
\frac{2M}{\hbar^{2}} \,   
E
\;  .
\label{eq:kappa}
\end{equation}
The regular boundary conditions at the origin (see the Appendix)
and at infinity lead to the selection of 
$J_{l} (\tilde k r)$ and $ K_{l} (\kappa r) $,
while the continuity of the logarithmic derivative  at  $r=a$
provides the eigenvalue equation
\begin{equation} 
\tilde k 
\,
\frac{J_l'(\tilde k  a)}{J_l(\tilde k  a)}
=
 \kappa 
\,
\frac{K_l'(\kappa  a)}{K_l(k a)}
\;  ,
\label{eq:eigenvalue_delta_circular_well}
\end{equation}
in which the primes denote derivatives.

The next step is the renormalization of the system. 
This is implemented by finding the
behavior of the running coupling constant from  the consistency requirement that 
the eigenvalue equation~(\ref{eq:eigenvalue_delta_circular_well})
admit a {\em finite\/} ground-state energy, when
  $a \rightarrow 0$.
As the analysis in Ref.~\cite{cam:dtII} shows, this system 
 can only sustain a bound state in the $s$ channel;
 this fact is confirmed by  Eq.~(\ref{eq:eigenvalue_delta_circular_well}), which
admits a nontrivial solution in the limit $a \rightarrow  0$
only for $l=0$.
Correspondingly,
the ground-state wave function becomes
\begin{equation}
 \Psi_{\! {\rm (gs)}}
  ({\bf r})=
\left\{\begin{array}{lr}
    B \; J_{0}  (\tilde k r)  
\;  & \textrm{for} 
\;\;    r < a  \;  ,
\\ 
   A\; K_{0}  (\kappa r)  
\;   & \textrm{for}  \;\;  r  > a  \;  ,
\\
  \end{array}\right. 
\;  
\label{eq:GS_2D_delta_circular_well}
\end{equation}
where the ratio between $ A $ and $ B$  can be determined from the additional continuity
condition
\begin{equation}
 B \,  J_{0}  ({\tilde k} a )
= A \, K_{0}  (\kappa a)
\; .
\label{eq:continuity_2D_delta_circular_well}
\end{equation}
For that particular channel ($l=0$),
Eqs.~(\ref{eq:wave_number_2D_delta_circular_well})
and (\ref{eq:eigenvalue_delta_circular_well}),
combined with the small-argument behavior of  Bessel functions
[in particular, 
Eq.~(\ref{eq:Mcdonald_0_small_arg})],
provide the desired running of the coupling
constant
\begin{equation}
\lambda (a)
\stackrel{(a \rightarrow 0)}{=}
- \frac{2 \pi}{ 
\left[ \ln \left( \kappa a/2 
\right)
+\gamma
\right]
}
\,
\left\{
1+ O \left(
\left[ \ln (\kappa a)
\right]^{-1}
\right)
\right\}
\;  .
\label{eq:delta_renormalized_coupling_circular_delta}
\end{equation}
In Eq.~(\ref{eq:delta_renormalized_coupling_circular_delta})
 the hierarchy of correction terms with respect to the variable 
\begin{equation}
\xi = \kappa a
\label{eq:xi}
\end{equation}
yields the three categories 
$ O( [\ln \xi ]^{-1}, \xi^{2} \ln \xi, \xi^{2} ) $,
including the corresponding higher orders;
of these terms, the first  is the dominant
one.
The order notation is used in Eq.~(\ref{eq:delta_renormalized_coupling_circular_delta})
and thereafter,
in order to keep track of all corrections with respect to
small arguments and regularizing parameters. This procedure is suggested by
the ill-defined nature of the formal 
expression~(\ref{eq:DELTA_time_rate_of_dilation_op_ddim-EV}),
which calls for 
a redefinition of each factor
before the limit $a \rightarrow 0$ is taken.
 However, it should be pointed out that the corresponding series are
typically going to be convergent rather than simply asymptotic,
despite our reference to asymptotic approximations.

For the computation of the anomaly~(\ref{eq:time_rate_of_dilation_op_ddim-EV}),
the values of the coefficients $A$ and $B$ should be determined.
First, they
are related by the
 condition~(\ref{eq:continuity_2D_delta_circular_well}),
 which  reduces to
\begin{equation}
B
\stackrel{(a \rightarrow 0)}{=}
-A \,
\left[ 
\ln \left( \frac{\kappa a}{2} \right) +\gamma 
\right] 
\,
\left\{
1 + O\left( 
\left[
\ln (\kappa a) 
\right]^{-1}
\right) 
\right\} 
\;  .
\label{eq:coeffs_2D_delta_circular_well}
\end{equation}
Secondly, their specific
 asymptotic values
  can be explicitly obtained from the
normalization condition 
\begin{eqnarray}
1
 & = & 
\int d^{2} {\bf  r} \, 
| 
\Psi_{\! {\rm (gs)}}
({\bf r})|^2  
 =
A^{2} \, 
2\pi
 \kappa^{-2} \, 
\left\{
{\mathcal K}(\kappa a) + 
\left( \frac{\kappa}{\tilde{k}} \right)^{2}
\,
\left(  \frac{B}{A}  \right)^{2}
\,
{\mathcal J}( \tilde{k}  a)  
\right\}
\nonumber \\
 & 
\stackrel{(a \rightarrow 0)}{=}
& 
\pi
 \kappa^{-2} \, 
A^{2} \, 
\left\{
1 + 
O
\left(
\kappa a  \, \ln [\kappa a] 
\right)
\right\}
\; ,
\label{eq:normaliz_circular_well_calculation}
\end{eqnarray}
where
\begin{equation}
{\mathcal K} (\xi)=
\int_{\xi}^{\infty} 
d z \, z
 \left[ K_{0} (z) \right]^{2} 
= \frac{1}{2} +
O
\left(
\xi \ln \xi  
\right)
\;
\label{eq:mathcal_K_integral}
\end{equation}
and
\begin{equation}
{\mathcal J} ( \tilde{\xi} )
=
\int_{0}^{  \tilde{\xi}  } 
d z \, z
 \left[ J_{0} (z) \right]^{2} 
= \frac{1}{2}\,  \tilde{\xi}^{2}
\left[ 1 +O(  \tilde{\xi} ^2) \right] 
\;  ,
\label{eq:mathcal_J_integral}
\end{equation}
with
\begin{equation}
\tilde{\xi}
 = 
\tilde{k} a
\;  .
\label{eq:xi_tilde}
\end{equation}
Thus, Eq.~(\ref{eq:normaliz_circular_well_calculation})
 implies that
\begin{equation}
A
\stackrel{(a \rightarrow 0)}{=}
\frac{\kappa}{\sqrt{\pi}}
\left\{
1 + 
O
\left(
\kappa a  \, \ln  [ \kappa a] 
\right)
\right\}
\;  .
\label{eq:Acoeff_2D_delta_circular_well}
\end{equation}

Now we proceed 
to  calculate the conformal anomaly via
the regularized version of Eq.~(\ref{eq:time_rate_of_dilation_op_ddim-EV}).
For the 
two-dimensional
$\delta$-function interaction,
this is accomplished by
using the regularized potential~(\ref{eq:delta_interaction_circular_well}),
in conjunction with the regularized wave function,
Eqs.~(\ref{eq:GS_2D_delta_circular_well}),
(\ref{eq:coeffs_2D_delta_circular_well}), and
(\ref{eq:Acoeff_2D_delta_circular_well}),
and running coupling~(\ref{eq:delta_renormalized_coupling_circular_delta}).
Then,
\begin{eqnarray}
\frac{d}{dt}
\left\langle D 
\right\rangle_{\scriptstyle \!  \Psi_{\rm \! {\scriptscriptstyle (gs)}}  }
 & = &
-\frac{1}{2}
\,
\int d^{2} {\bf r}
\,
V ({\bf r})
\,
{\mathcal E}_{\bf r}
\left|
\Psi_{{\rm (gs)}} ({\bf r})
\right|^{2}
\;
  \nonumber \\
& = & 
\frac{\hbar^{2}}{2M }
\frac{ 2 \lambda \, B^{2} }{ \tilde{ \xi}^{2}}
\,
\int_{0}^{ \tilde{\xi}}
d z \, z
\,
J_{0} (z)
\,
{\mathcal E}_{ z}
J_{0} (z)
\;
  \nonumber \\
&  
\stackrel{(a \rightarrow 0)}{=}
&
E_{{\rm (gs)}} 
\,
\left\{
1 + O\left( 
\left[
\ln (\kappa a) 
\right]^{-1}
\right) 
\right\} 
\;  ,
\label{eq:anomaly_2D_delta_circular_well}
\end{eqnarray}
as follows from
 Eqs.~(\ref{eq:GS_energy}) and (\ref{eq:wave_number_2D_delta_circular_well}), 
as well as from
from the small-argument behavior
of $J_{0}(z)$;
in Eq.~(\ref{eq:anomaly_2D_delta_circular_well}),
${\mathcal E}_{ z}$ is the one-dimensional (radial) generalization
of Eq.~(\ref{eq:Eulerian_derivative}).
Finally, taking the limit
 $a \rightarrow 0$,
the conformal anomaly~(\ref{eq:anomaly_2D_delta_circular_well}) 
is in perfect agreement with the
expected answer~(\ref{eq:time_rate_of_dilation_op-EV}).

\subsection{Real-space 
regularization with a radial 
$\mbox{\boldmath $\delta$}$-function 
interaction}
\label{sec:circular_delta}

An alternative
real-space regularization 
technique
is provided by a radial 
$\delta$-function interaction
\begin{equation}
V({\bf r})  \sim 
g
\,
\frac{ \delta (r-a)}{2 \pi \alpha \, a}
\equiv
- 
\frac{ \hbar^{2}}{2M} 
\, \hat{\lambda}
\,
\frac{ \delta (r-a)}{2 \pi a}
\;  .
\label{eq:delta_interaction_radial_delta}
\end{equation}
In addition to defining 
the regularized potential,
Eq.~(\ref{eq:delta_interaction_radial_delta})
introduces two 
auxiliary quantities:
(i)
an arbitrary proportionality factor $\alpha$
associated with
a possible ambiguity in the definition of the radial 
$\delta$ function~\cite{radial_delta};
(ii) 
a reduced coupling $\hat{\lambda}= \lambda/\alpha$.
As it is to be expected, the anomaly~(\ref{eq:DELTA_time_rate_of_dilation_op_ddim-EV}), 
to be computed later in this section, will be 
independent of the  undetermined
``ambiguity factor'' $\alpha$.

Due to the central nature of Eq.~(\ref{eq:delta_interaction_radial_delta}),
the formalism of the Appendix
can be directly applied again.
The corresponding
 Schr\"odinger equation 
for the reduced radial wave function $u_{l}(r)$ 
is now given by
 \begin{equation}
\left[ 
 \frac{d^2}{dr^2}
+ 
\frac{2M \, E }{\hbar^{2}}  +
\frac{\hat{\lambda} }{2 \pi a}
\,
\delta (r-a)
-
\frac{l^{2} -1/4 }{r^2}
 \right]
u_{l} (r)
=
0
\;  .
\label{eq:radial_Schr_2D_delta_radial_delta}
\end{equation}
The bound-state solution ($E<0$)
to Eq.~(\ref{eq:radial_Schr_2D_delta_radial_delta}),
subject to the regular boundary conditions at the origin (see the Appendix)
and at infinity,
is given by
\begin{equation}
  R_{l}(r)
\equiv
\frac{ u_{l}(r) }{\sqrt{r}}
=\left\{\begin{array}{lr}
B_{l}
\,
 I_{l} (\kappa r) 
\;  
& \textrm{for}
\;\; r<a  \; ,
\\
A_{l}
\,
 K_{l} (\kappa r) 
\;  
 & \textrm{for} \;\; r>a  \; , \\
  \end{array}
\right.
\;   \;   
\label{eq:Bessel_solution_2D_delta_radial_delta}
\end{equation}
where
$ \kappa$
is defined just as in Sec.~\ref{sec:circular_well},
Eq.~(\ref{eq:kappa}).

The
 eigenvalue equation
follows from the condition defining the $\delta$-function
discontinuity at $r=a$,
\begin{equation}
\left. 
\frac{d u_{l}}{dr}
\right|_{r=a^{+}}
-
\left. 
\frac{d u_{l}}{dr}
\right|_{r=a^{-}}
=
-
\frac{\hat{\lambda} }{2 \pi a}
\,
u_{l}(a)
\;  .
\label{eq:delta-function_discontinuity}
\end{equation}
Therefore, 
with the functions defined in Eq.~(\ref{eq:Bessel_solution_2D_delta_radial_delta}),
the eigenvalue equation takes the explicit form
\begin{equation} 
\kappa
\left[
A_{l}  \,
K_{l}'(\kappa a)
-
B_{l}  \,
I_{l}'(\kappa a)
\right]
=
-
\frac{\hat{\lambda} }{2 \pi a}
A_{l} 
\,
K_{l}(\kappa a)
\;  ,
\label{eq:eigenvalue_delta_radial_delta}
\end{equation}
where the primes denote derivatives.
A detailed analysis of 
Eq.~(\ref{eq:eigenvalue_delta_radial_delta})
shows, as in Sec.~\ref{sec:circular_well},
that a nontrivial solution exists only for $l=0$.
Correspondingly,
the ground-state wave function becomes
\begin{equation}
 \Psi_{\! {\rm (gs)}}
   ({\bf r})=
\left\{\begin{array}{lr}
    B \; I_{0}  (\kappa r)    
\;  & \textrm{for} 
\;\;    r < a  \;  , \\ 
   A\; K_{0}  (\kappa r)  
\;   & \textrm{for}  \;\;  r  > a  \; ,
\\
  \end{array}\right. 
\;  
\label{eq:GS_2D_delta_radial_delta}
\end{equation}
where $A \equiv A_{0}$ and
 $B \equiv B_{0}$.
In addition, 
 $ A $ and $ B$  can be determined from the continuity
condition
\begin{equation}
 B \,  I_{0} (\kappa a)
= A \, K_{0}(\kappa a)
\;  ,
\label{eq:continuity_2D_delta_radial_delta}
\end{equation}
 which 
reduces to
\begin{equation}
B
\stackrel{(a \rightarrow 0)}{=}
-A \,
\left[ 
\ln \left( \frac{\kappa a}{2} \right) +\gamma 
\right] 
\,
\left\{
1 + O \left( 
 \left[ \kappa a \right]^{2} 
\right)
\right\} 
\;  .
\label{eq:coeffs_2D_delta_radial_delta}
\end{equation}
Moreover, 
the
normalization condition  gives
\begin{equation}
1
  = 
\int d^{2} {\bf r} \, 
| \Psi_{\! {\rm (gs)}}
({\bf r})|^2  
 =
A^{2} \, 
2\pi
 \kappa^{-2} \, 
\left\{
{\mathcal K}(\kappa a) + 
\left(  \frac{B}{A}  \right)^{2}
\,
{\mathcal I}( \kappa  a)  
\right\}
\; ,
\label{eq:normaliz_radial_delta_calculation}
\end{equation}
where
${\mathcal K} (\xi)$ is defined in 
Eq.~(\ref{eq:mathcal_K_integral}), while
\begin{equation}
{\mathcal I} ( \xi )
=
\int_{0}^{  \xi  } 
d z \, z
 \left[ I_{0} (z) \right]^{2} 
= \frac{1}{2}\,  \xi^{2}
\left[ 1 +O(  \xi^2) \right] 
\;  .
\label{eq:mathcal_I_integral}
\end{equation}
Thus,
Eqs.~(\ref{eq:mathcal_K_integral}), (\ref{eq:normaliz_radial_delta_calculation}), and
(\ref{eq:mathcal_I_integral})
lead again to an expression identical to
Eq.~(\ref{eq:Acoeff_2D_delta_circular_well}).

The running of the coupling constant is obtained 
by replacing the small-argument behavior of  Bessel functions
[in particular, 
Eq.~(\ref{eq:Mcdonald_0_small_arg})]
in 
Eq.~(\ref{eq:eigenvalue_delta_radial_delta}). 
Then,
\begin{equation}
\hat{\lambda} (a)
\equiv
\frac{ \lambda (a)}{  \alpha}
\stackrel{(a \rightarrow 0)}{=}
- \frac{2 \pi}{ 
\left[ \ln \left( \kappa a/2 
\right)
+\gamma
\right]
}
\,
\left\{
1+ O \left(
\left[ \kappa a  \right]^{2} 
\ln [\kappa a]
\right)
\right\}
\;  ,
\label{eq:delta_renormalized_coupling_radial_delta}
\end{equation}
where the correction terms with respect to the variable $\xi = \kappa a$ 
appear in the two categories
$ O( \xi^{2} \ln \xi, \xi^{2} ) $,
including the corresponding higher orders;
of these terms, the first  is the dominant
one.

Finally,
the conformal anomaly can be computed by replacing
the regularized potential~(\ref{eq:delta_interaction_radial_delta}),
 the regularized wave function
[Eqs.~(\ref{eq:Acoeff_2D_delta_circular_well}),
(\ref{eq:GS_2D_delta_radial_delta}), and
(\ref{eq:coeffs_2D_delta_radial_delta})],
and
the running coupling~(\ref{eq:delta_renormalized_coupling_radial_delta})
in Eq.~(\ref{eq:time_rate_of_dilation_op_ddim-EV}).
This computation yields
\begin{eqnarray}
\frac{d}{dt}
\left\langle D 
\right\rangle_{\scriptstyle \!  \Psi_{\rm \! {\scriptscriptstyle (gs)}}  } 
 & = &
-\frac{1}{2}
\,
\int d^{2} {\bf r}
\,
V ({\bf r})
\,
{\mathcal E}_{\bf r}
\left|
\Psi_{\! {\rm (gs)}}
 ({\bf r})
\right|^{2}
\;
  \nonumber \\
& = & 
\frac{\hbar^{2}}{2M }
\frac{  \hat{\lambda}}{2}
\,
\int_{0}^{\infty}
 d r
\,
\delta (r-a)
\,
{\mathcal E}_{\bf r}
\left|
\Psi_{\! {\rm (gs)}}
 ({\bf r})
\right|^{2}
\;  .
\label{eq:anomaly_2D_delta_radial_delta_prelim}
\end{eqnarray}
As the wave function~(\ref{eq:GS_2D_delta_radial_delta})
has a  discontinuous derivative through $r=a$,
the integral in Eq.~(\ref{eq:anomaly_2D_delta_radial_delta_prelim})
is conveniently computed by dividing the interval 
$[0,\infty)$ into the subintervals  
$I_{<}=[0,a]$ and $I_{>}=[a,\infty)$.
Then,
\begin{equation}
\frac{d}{dt}
\left\langle D 
\right\rangle_{\scriptstyle \!  \Psi_{\rm \! {\scriptscriptstyle (gs)}}  } 
  = 
{\mathcal A}^{(<)} 
+
{\mathcal A}^{(>)} 
\;  ,
\label{eq:anomaly_2D_delta_radial_delta_prelim2}
\end{equation}
where 
\begin{equation}
{\mathcal A}^{(j)} 
=
\frac{\hbar^{2}}{2M }
\frac{  \hat{\lambda}}{2}
\,
\int_{I_{j}}
dr \,
\delta (r-a)
\,
{\mathcal E}_{\bf r}
\left|
\Psi^{(j)}_{{\rm (gs)}} ({\bf r})
\right|^{2}
=
\frac{\hbar^{2}}{2M }
\frac{  \hat{\lambda}}{2}
\,
R^{(j)} (a)
\,
{\mathcal E}_{ r}
R^{(j)} (a)
\;  ,
\end{equation}
with 
 $j=<$ when $r<a$ 
and
 $j=>$ when $r>a$,
while
$\Psi^{(j)}_{{\rm (gs)}} ({\bf r}) \equiv
R^{(j)} (r)$ is 
given in Eq.~(\ref{eq:GS_2D_delta_radial_delta}).
From
 Eqs.~(\ref{eq:Acoeff_2D_delta_circular_well})
and
(\ref{eq:GS_2D_delta_radial_delta}), 
as well as the small-argument
behavior of Bessel functions,
the exterior integral becomes 
\begin{equation}
{\mathcal A}^{(>)} 
=
\frac{\hbar^{2}}{2M }
\frac{  \hat{\lambda} \, A^{2} }{2}
\,
\xi
\, 
K_{0} ( \xi)
\,
K'_{0} ( \xi)
\stackrel{(a \rightarrow 0)}{=}
E_{{\rm (gs)}} 
\,
\left\{
1 + O \left( \xi^{2} \ln \xi
\right)
\right\} 
\;  
\end{equation}
[where $\xi$ is defined in Eq.~(\ref{eq:xi})],
while the
interior integral
takes the form
\begin{equation}
{\mathcal A}^{(<)} 
=
\frac{\hbar^{2}}{2M }
\frac{  \hat{\lambda} B^{2} 
 }{2}
\,
 \xi
\,
I_{0} ( \xi)
\,
I'_{0} ( \xi)
\stackrel{(a \rightarrow 0)}{=}
E_{{\rm (gs)}} 
\,
\times
 O\left( \xi^{2} \ln \xi
\right) 
\;  ;
\end{equation}
as a result,
\begin{equation}
\frac{d}{dt}
\left\langle D
\right\rangle_{\scriptstyle \!  \Psi_{\rm \! {\scriptscriptstyle (gs)}}  } 
\stackrel{(a \rightarrow 0)}{=}
E_{{\rm (gs)}} 
\,
\left\{
1 + O\left( 
\left[ \kappa a \right]^{2}
\ln  [ \kappa a ] 
\right) 
\right\} 
\;  .
\label{eq:anomaly_2D_delta_radial_delta}
\end{equation}
Therefore,
when the limit
 $a \rightarrow 0$ is enforced,
the conformal anomaly~(\ref{eq:anomaly_2D_delta_circular_well}) 
again agrees with the predicted value, Eq.~(\ref{eq:time_rate_of_dilation_op-EV}).

\subsection{Dimensional regularization}
\label{sec:dimensional_regularization}

In dimensional regularization~\cite{bol:72}
the modification of the short-distance physics is nontrivially
accounted for
by a dimensional generalization of 
the theory---the relevant physics is analytically continued 
 from a given physical dimensionality
$d_{0}$ to $d=d_{0}-\epsilon$, with $\epsilon =0^{+}$.
For singular interactions,
 this procedure is
implemented by properly extending the potential 
from $d_{0}$ to $d$ dimensions.
Even though this generalization is somewhat arbitrary, 
in this paper we follow the convenient prescription  provided in
Refs.~\cite{cam:dtI,cam:dtII,pi_singular,pi_delta}.
Accordingly,
\begin{equation}
V({\bf r})  \sim 
g
 \, \mu^{\epsilon} \,
\delta^{({d})} ({\bf r})
\equiv
- 
\frac{ \hbar^{2}}{2M} 
\lambda
 \, \mu^{\epsilon} \,
\delta^{({d})} ({\bf r})
\;  ,
\label{eq:delta_interaction_DR}
\end{equation}
where
the physical dimensions of the original theory are preserved
by changing the dimensions of the
coupling according to
$
g
\rightarrow 
g
\, \mu^{\epsilon}
$~\cite{cam:dtI}.

The interaction~(\ref{eq:delta_interaction_DR})
can be regarded as effectively central, so that the 
results of the 
Appendix
can be applied.
The corresponding
 Schr\"odinger equation 
for the reduced radial wave function $u_{l}(r)$,
in $d \equiv 2 \nu + 2 =2-\epsilon$ dimensions, 
is given by 
\begin{equation}
\left[ 
\frac{d^{2}}{dr^{2}}
+
\frac{2M  E }{\hbar^{2}} 
 -  
\frac{(l + \nu)^{2}
 - 1/4 }{r^{2}}
\right]  
u_{l}(r) = 0
\;  ,
\label{eq:radial_Schr_2D_delta_DR}
\end{equation}
for $r \neq 0$.
Equation~(\ref{eq:radial_Schr_2D_delta_DR})
is formally identical to that
of a free particle, but it is to be supplemented
by the stringent boundary condition
enforced by the $\delta$-function singularity at the origin.

The bound-state solution ($E<0$)
to Eq.~(\ref{eq:radial_Schr_2D_delta_DR})
is a linear combination of the Bessel functions $ I_{l+\nu}(\kappa r) $ 
and $ K_{l+\nu} (\kappa r)$. 
As usual,
the boundary condition at infinity  leads to the rejection of 
$I_{l+\nu}(\kappa r) $. As for
the modified Bessel function $K_{l+\nu}(\kappa r) $, 
its small-argument behavior 
for $l \neq 0$ leads to  a singular term 
proportional to $r^{-(l+\nu)}$. Thus, the 
boundary condition at the origin can {\em only\/} 
be satisfied for $ l=0 $, 
a result that agrees with the conclusions
drawn from real-space regularization techniques.
As a consequence, the regularized 
radial wave
function is of the form
$ R (r)  \propto r^{-\nu} \, K_{\nu} (\kappa r)$,
and the corresponding
 normalized ground-state wave function becomes
\begin{equation}
 \Psi_{\! {\rm (gs)}}
({\bf r})
= 
\frac{\kappa}{\sqrt{\pi}}
\,
\frac{ \pi^{\epsilon/4}} {[\Gamma (1+ \epsilon/2) ]^{1/2} }
 \,
r^{\epsilon/2 }  
\,
K_{-\epsilon/2} (\kappa r)
\equiv  A (\epsilon) \,
F_{\epsilon} (\kappa r)
\;  ,
\label{eq:GS_2D_delta_DR}
\end{equation}
with
\begin{equation}
F_{\epsilon} (z)
\equiv z^{\epsilon/2} K_{-\epsilon/2} (z)
\; .
\label{eq:aux_F_Bessel}
\end{equation}
In Eq.~(\ref{eq:GS_2D_delta_DR})
the normalization constant 
$A (\epsilon)$
in $d= 2(\nu +1)=
2-\epsilon$ dimensions was obtained 
from $\int_{0}^{\infty} dz \, z \left[ K_{p}(z) \right]^{2}=
\Gamma (1+p ) \, \Gamma (1-p)/2$, 
and its asymptotic value is 
\begin{equation}
A (\epsilon)
\stackrel{(\epsilon \rightarrow 0)}{=}
 \frac{\kappa}{\sqrt{\pi} } \,  [1+O(\epsilon)]
\; .
\end{equation}

The fact that Eq.~(\ref{eq:GS_2D_delta_DR}) is the only bound state
is a requirement of the eigenvalue equation,
which 
follows by asking that the
$\delta$-function singularity at the origin be enforced~\cite{cam:dtI}. 
This procedure implies the condition
\begin{equation}
\frac{\lambda \, \mu^{\epsilon}}{4\pi}
\left(  \frac{2M}{\hbar^{2}} \,
\frac{|E|}{4\pi}
\right)^{-\epsilon/2}
\Gamma
\left(\frac{\epsilon}{2} \right) = 1
\; ,
\label{eq:delta_eigen2}
\end{equation}
which displays a simple pole at $\epsilon=0$, making the theory singular for 
the two-dimensional unregularized case.
However, for  $\epsilon=0^{+}$, 
Eq.~(\ref{eq:delta_eigen2})
permits the existence of the bound state~(\ref{eq:GS_2D_delta_DR}).

Renormalization is implemented
by introducing the running
coupling~\cite{cam:dtI,cam:dtII,pi_delta},
which is determined in the limit $\epsilon \rightarrow 0$ from
Eq.~(\ref{eq:delta_eigen2}), i.e.,
\begin{equation}
\lambda(\epsilon) 
\stackrel{(\epsilon \rightarrow 0)}{=}
2 \pi \epsilon
\left\{ 1 + \frac{\epsilon}{2}  \,
\left[
g^{(0)}
- \left(  \ln 4\pi - \gamma \right)
\right]
+ o (\epsilon)
\right\}
\; ,
\label{eq:delta_renormalized_coupling_DR}
\end{equation}
where $\gamma$ is the Euler-Mascheroni constant
and $g^{(0)}$ is an arbitrary finite part. In particular, 
from Eq.~(\ref{eq:delta_renormalized_coupling_DR}),  the ground-state 
energy becomes
\begin{equation}
E_{_{\rm (gs)}}
=
- 
\frac{\hbar^{2}\mu^{2}}{2M}
\, 
e^{ g^{(0)}  }
\;  .
\label{eq:delta_gs_energy_DR}
\end{equation}

Finally,
the conformal anomaly can be computed
by replacing
the regularized potential~(\ref{eq:delta_interaction_DR})
and the running coupling~(\ref{eq:delta_renormalized_coupling_DR})
in
Eq.~(\ref{eq:time_rate_of_dilation_op_ddim-EV}).
Then,
\begin{eqnarray}
\frac{d}{dt}
\left\langle
D
\right\rangle_{\scriptstyle \!  \Psi_{\rm \! {\scriptscriptstyle (gs)}}  } 
 &  =   &
\left[
- \frac{ \hbar^{2}}{2M} \, \lambda (\epsilon) \mu^{\epsilon} 
\right]
\,
\left\{
\frac{\epsilon}{2}
\left\langle
  \delta^{(d)}({\bf r}) 
\right\rangle_{\scriptstyle \!  \Psi_{\rm \! {\scriptscriptstyle (gs)}}  } 
-
\frac{1}{2}
\int d^{d} {\bf r}
 \, 
\delta^{(d)} ({\bf r}) 
\, 
{\mathcal E}_{\bf r} 
\left| 
 \Psi_{\! {\rm (gs)}}
   ({\bf r})
\right|^{2}   
\right\}
 \nonumber   \\
&      \stackrel{(\epsilon \rightarrow 0)}{=}
  &  
\left[
- \frac{ \hbar^{2}}{2M} 
\right]
\,  \pi \epsilon
\,
\left\{
\epsilon 
\,
\left| 
 \Psi_{\! {\rm (gs)}}
   ({\bf 0})
\right|^{2}   
-
\left[
{\mathcal E}_{\bf r} 
\left| 
 \Psi_{\! {\rm (gs)}}
   ({\bf r})
\right|^{2}   
\right]_{{\bf r}= {\bf 0} }
\right\}
\,
\left[ 1 +O(\epsilon) 
\right]
 \nonumber   \\
&      \stackrel{(\epsilon \rightarrow 0)}{=}
  &  
E_{{\rm (gs)}} 
\, 
\left\{
\left[
\epsilon 
\,
F_{\epsilon} (0)
\right]^{2}
-
2 \epsilon 
\,
F_{\epsilon} (0)
\lim_{z \rightarrow 0}
{\mathcal E}_{z} 
F_{\epsilon} (z)
\right\}
\,
\left[
1 +
O(\epsilon)
\right]
\label{dr1}
\; ,
\end{eqnarray}
where Eqs.~(\ref{eq:GS_energy})
and (\ref{eq:GS_2D_delta_DR}) were applied
 in the final line,
while
${\mathcal E}_{ z}$ is the one-dimensional (radial) generalization
of Eq.~(\ref{eq:Eulerian_derivative}).
The operations to be performed 
in Eq.~(\ref{dr1}) at the level of
the regularized wave function~(\ref{eq:aux_F_Bessel})
can be simplified 
with the use of Bessel-function identities.
First, 
the  small-argument expansion 
\begin{equation}
 K_{p}(z) 
       \stackrel{(z \rightarrow 0)}{=}  
\frac{1}{2} 
\,
\left[
\Gamma (p) 
\left(
\frac{z}{2}
 \right)^{-p}
+
\Gamma (-p) 
\left(
\frac{z}{2}
 \right)^{p}
\right]
\,
\left[
1 +
O \left(z^{2} \right)
\right]
\; ,
\label{eq:Macdonald_small_argument}
\end{equation}
leads to
$F_{\epsilon}(0) =1/\epsilon$,
with corrections of order $O(\epsilon)$.
Secondly, either from Eq.~(\ref{eq:Macdonald_small_argument}) again or 
from the
identity
\begin{equation}
\frac{1}{z}
\,
\frac{d }{d z}
\left[
z^{-p} 
K_{p} (z) 
\right]
= - z^{-(p+1)}
\, K_{p+1}(z)
\;  ,
\end{equation}
one concludes that
$
{\mathcal E}_{\xi} 
 F_{\epsilon}(\xi) =-\xi^{\epsilon}$,
with corrections of orders $O(\epsilon, \xi^{2})$.
Therefore,
the final result is
\begin{eqnarray}
\frac{d}{dt} 
\left\langle D 
\right\rangle_{\scriptstyle \!  \Psi_{\rm \! {\scriptscriptstyle (gs)}}  }
&        \stackrel{(\epsilon \rightarrow 0)}{=}  & 
E_{{\rm (gs)}} 
\, 
\left\{
\left[
1 +
O(\epsilon)
\right]
+
2 \,
\lim_{r \rightarrow 0}
\left(  \kappa r \right)^{\epsilon}
\left[
1 +
O(\epsilon)
\right]
\right\}
\nonumber \\
 &     \stackrel{(\epsilon \rightarrow 0)}{=} &
E_{{\rm (gs)}} 
\, 
\left[
1 +
O(\epsilon)
\right]
\;  .
\label{eq:anomaly_2D_delta_DR}
\end{eqnarray}
Remarkably,
the limit
 $\epsilon \rightarrow 0$ should be
taken only as the last step, as required by the dimensional-regularization 
prescription.
When this procedure is properly applied,
the conformal anomaly~(\ref{eq:anomaly_2D_delta_DR}) 
again agrees with the 
value
  anticipated in
Eq.~(\ref{eq:time_rate_of_dilation_op-EV}).

\section{Scattering Sector for a Two-Dimensional
$\mbox{\boldmath $\delta$}$-Function 
Interaction}
\label{sec:scattering}

In Sec.~\ref{sec:2D_delta} we focused our analysis on the
emergence of extra terms for relevant
bound-state expectation values.
However,
a complete characterization of the interaction requires
the complementary  analysis in the scattering sector of the theory.
For the two-dimensional 
$\delta$-function interaction,
the compatibility of the renormalization in both sectors is well known; e.g.,
as discussed in Refs.~\cite{jac:91,cam:dtI,cam:dtII,pi_delta}.
In this section
we now complete our analysis by using these compatibility
requirements and consequently display the emergence of an anomalous commutator algebra 
for scattering.

Our goal is to make use of 
 expectation values and thereby construct
nonvanishing symmetry-breaking terms.
Unfortunately, this construction proves to be considerably more difficult than for 
bound states because of the nonexistence of scattering states that are simultaneously
normalized and stationary.
The proper formalism to display the anomalous terms is then provided 
by  time-dependent  collision theory~\cite{gottfried}.
In addition, 
we conveniently switch to the Schr\"{o}dinger picture
as the natural way to study the time evolution of these wave packets.

Let $\Psi ({\bf r},t)$
be a wave packet of positive energy 
evolving in $d$ dimensions from 
an
initial state 
\begin{equation}
\Psi ({\bf r},0) =
\int \frac{ d^{d} {\bf q}}{ (2 \pi)^{d}}
\,
\chi ({\bf q})
\,
\phi_{ {\bf q}} ({\bf r})
\;  ,
\label{eq:initial_wave_packet}
\end{equation}
in which $\phi_{ {\bf q}} ({\bf r}) = \exp \left( i
 {\bf q} \cdot {\bf r} \right)$.
This function $\phi_{ {\bf q}} ({\bf r}) $ can be interpreted as the incident
  plane wave for a scattering experiment in which
\begin{equation}
\psi_{ {\bf q}} ({\bf r})
=
\phi_{ {\bf q}} ({\bf r})
+
\frac{2M}{\hbar^{2}}
\int  d^{d} {\bf r'}
\,
{\mathcal G}^{(+)} _{d} ({\bf r}-{\bf r'}; q)
\,
V({\bf r'})
\,
\psi_{ {\bf q}} ({\bf r'})
\;
\label{eq:L_S}
\end{equation}
is the stationary-state interacting wave function
for energy $E= E_{{\bf q}} 
\equiv 
\hbar^{2}q^{2}/2M$ and
 $q=|{\bf q}|$, while
${\mathcal G}^{(+)} _{d} ({\bf R}; q)
=
-
i \,
\left( q/2\pi R \right)^{d/2-1}
H^{(1)}_{d/2-1}(qR)/4$  
stands for the corresponding causal Green's function.

In this section we assume that the applicability 
conditions and approximations of Ref.~\cite{gottfried}
are satisfied for the treatment with wave packets.
In this context,
the present derivation is at least sufficient to 
prove our claim of the existence of
 anomalous terms in well-defined expectation values.
Then, 
starting with the initial condition~(\ref{eq:initial_wave_packet}),
the time evolution of the state
$ \left| \Psi (t)  \right\rangle $
is asymptotically described by
\begin{equation}
\Psi ({\bf r},t) =
\int \frac{ d^{d} {\bf q}}{ (2 \pi)^{d}}
\,
\chi ({\bf q})
\psi_{ {\bf q}} ({\bf r})
\,
e^{-i \omega_{{\bf q}} t}
\;  ,
\label{eq:evolution_wave_packet}
\end{equation}
where $\omega_{{\bf q}} = E_{{\bf q}}/\hbar $.
In general, using a resolution of the form~(\ref{eq:evolution_wave_packet})
between states 
$ \left| \Psi_{1} (t)  \right\rangle $
and
$ \left| \Psi_{2} (t)  \right\rangle $,
the  expression for the 
 transition matrix elements of an operator $A$ becomes
\begin{equation}
 \left\langle 
\Psi_1 (t) 
\left| A \right| 
\Psi_2  (t) 
\right\rangle =
 \int 
\frac{d^d  {\bf q''} }{(2\pi)^d} 
\int 
\frac{d^d  {\bf q' } }{(2\pi)^d} 
\,
\chi_{1}^{*} ({\bf q''})    
\chi_{2} ({\bf q'})
\,
  e^{-i( \omega_{\bf q'} -\omega_{\bf q''}) \, t }  
\,
\left\langle \psi_{{\bf q}''} 
\left| A \right| 
\psi_{{\bf q'}}  \right\rangle
\;  .
\label{eq:transition_elements}
\end{equation}
In particular, for the expectation value of the dilation operator,
\begin{equation}
\frac{d}{dt}
\left\langle
D
\right\rangle_{\scriptstyle \!  \Psi (t) }
=
\int \frac{d^d  {\bf q''} }{(2\pi)^d}
 \int \frac{d^d  {\bf q'}  }{(2\pi)^d}    
\,
\chi^{*} ({\bf q''})       \chi({\bf q'}) 
\,
    e^{-i( \omega_{\bf q'} 
-\omega_{\bf q''})
t}
\left\langle 
\psi_{{\bf q}''} 
\left | 
\frac{ [D,H]_{\rm extra} }{ i \hbar }
\right|  
\psi_{{\bf q'}}  \right\rangle
\;  ,
\label{eq:t_evolution_EV_dilation_prelim}
\end{equation}
an expression which can be evaluated for a  specific two-dimensional
potential $V({\bf r})$
from Eq.~(\ref{eq:time_rate_of_dilation_op_ddim}), whence
 \begin{eqnarray}
  \left\langle 
\psi_1 \left| 
\frac{ [D,H]_{\rm extra} }{i \hbar} 
\right| 
\psi_2  
\right\rangle
  & =&  
\frac{1}{2}
 \int d^{2} {\bf r}  
\,  \psi^{*}_{1} 
({\bf r}) 
\,   
 {\bf \nabla} \cdot 
\,   
 ({\bf r} V({\bf r})  ) 
 \psi_2 ({\bf r})
  \nonumber \\
 & = & 
-\frac{1}{2} 
\int d^2 {\bf r}
\,   
V({\bf r})
\,
{\mathcal E}_{\bf r} 
\left[
\psi_{1}^{*}({\bf r}) \psi_{2} ({\bf r})
\right]
\;   .
 \end{eqnarray}
Therefore, for the two-dimensional 
$\delta$-function interaction,
\begin{eqnarray}
   \left\langle 
\psi_{{\bf q}''} 
\left | 
\frac{ [D,H]_{\rm extra} }{i \hbar} 
\right| \psi_{{\bf q}'}  
\right\rangle
 & = & 
-\frac{g}{2} 
\left\lbrace 
\psi_{{\bf q}''}^*(0)  
\left[
\hat{\mathcal E}_{{\bf r}} 
\psi_{{\bf q}'}
({\bf r})
\right]_{{\bf r}=0}
 + \left[
\hat{\mathcal E}_{{\bf r}} 
\psi_{{\bf q}''}^*({\bf r})  
\right]_{{\bf r}=0} 
\psi_{{\bf q}'}(0)  
\right\rbrace
\nonumber \\
& = &
 \left( -\frac{\hbar^2}{2 M} 
\right) 
\,
4 \sqrt{q'' q'}
\,
  f_{q''}^*({\bf \Omega}^{(2)})  
\,
f_{q'} ({\bf \Omega}^{(2)})  
\;  ,
\label{eq:comm_D_H}
\end{eqnarray}
where 
\begin{equation}
f_{q}
\equiv
f_{q}
({\bf \Omega}^{(2)})  
=
\sqrt{\frac{2 \pi}{q}}
\,
\left[ \ln  \left(  
\frac{E_{\bf q}}{ | E_{{\rm (gs)}} |   } 
\right)  
-i \,\pi 
\right]^{-1}       
\; 
\end{equation}
is the isotropic two-dimensional scattering amplitude~\cite{pi_delta},
in which the angular dependence  
${\bf \Omega}^{(2)}  $ 
can be omitted.
Finally, from 
Eqs.~(\ref{eq:t_evolution_EV_dilation_prelim})
and (\ref{eq:comm_D_H}),
\begin{equation}
\frac{d}{dt}
\left\langle
D
\right\rangle_{\scriptstyle \!  \Psi (t) }
=
\left(-\frac{2\hbar^2}{M}   \right)  
 \int 
\frac{d^{2}  {\bf q''}}{(2\pi)^{2} }
 \int \frac{d^{2}  {\bf q'}}{(2\pi)^{2} }    
\,
\chi^{*}({\bf q''})    \, 
\chi({\bf q'})   
\,
  e^{-i( \omega_{\bf q'} -\omega_{\bf q''}) \, t}
 \,
 \sqrt{q'' q'} 
\,
f_{q''}^{*} ({\bf \Omega}^{(2)})   
\,
f_{q'} ({\bf \Omega}^{(2)})  
\;  ,
\end{equation}
whence it 
follows that
\begin{equation}
\frac{d}{dt}
\left\langle 
D 
\right\rangle_{\scriptstyle \!  \Psi (t) }
=
-
\frac{\hbar^{2}}{2M}
\,
\left| 
{\mathcal F} (t)
 \right|^{2}
\;  ,
\label{eq:time_rate_of_dilation_op-EV_scatt}
\end{equation}
where
\begin{equation}
  {\mathcal F} (t) 
=
2 \int 
 \frac{ d^{2} {\bf q}}{ (2 \pi)^{2}}
\,
\chi ({\bf q}) 
e^{-i \omega_{{\bf q}} t}
\sqrt{q}
\,
f_{q}
\;  .
\label{eq:time_rate_of_dilation_op-EV_scatt2}
\end{equation}

Equations~(\ref{eq:time_rate_of_dilation_op-EV_scatt}) and
(\ref{eq:time_rate_of_dilation_op-EV_scatt2}) can be verified in the three regularization
schemes introduced earlier
and constitute the main result of this section.
Just as the logarithmic behavior of Eqs.~(\ref{eq:delta_wf_normalized_renormalized})
and (\ref{eq:Mcdonald_0_small_arg}) was the source of the nontrivial anomalous commutator 
in the bound state sector,
the singular behavior
$H^{(1)}_{0}(z)  
\stackrel{(z \rightarrow 0)}{=}
 1 + 2 i  \left[ \ln (z/2) + \gamma \right]/\pi
$
[up to terms $O(z^{2})$]
in
Eq.~(\ref{eq:L_S})
yields the nontrivial 
expression~(\ref{eq:time_rate_of_dilation_op-EV_scatt}).
This final result admits the following interpretation:
a wave packet
undergoes
a time evolution dictated by the linear superposition of its initial Fourier components;
as the scattering amplitude depends upon a scale $|E_{(gs)}|$, the 
ensuing symmetry charges are 
no longer conserved.

\section{Conclusions}
\label{sec:conclusions}

In conclusion, we have explicitly shown 
the anomalous nature of the 
commutator algebra 
in conformal quantum mechanics
for the two-dimensional
$\delta$-function interaction.
These results are supported by detailed computations performed with three distinct
regularization techniques in both the bound-state sector and the scattering sector of the 
interacting theory.

The crucial point in this anomalous behavior 
is that extra terms 
in the  commutators of the SO(2,1) generators arise from the dimensionally-transmuted scale
of the renormalized theory.
The  implication of the existence of these nonvanishing terms at the level of nonconserved
symmetry charges
was explored and general
properties of quantum-mechanical averages were used to shed light on the
physical meaning of our results.

A similar  but considerably subtler analysis 
can be applied to the inverse square potential
in any number of dimensions;
the details of this procedure
will be discussed elsewhere.
Incidentally, an 
alternative technique that has been widely used to deal with singular
potentials is the method of self-adjoint extensions~\cite{jac:91}.
It would be interesting to investigate the same issues 
using that method and to provide a comparison
with our results~\cite{esteve}.

\acknowledgments{This research was supported in part by
an Advanced Research Grant from the Texas
Higher Education Coordinating Board 
and by the University of San Francisco Faculty Development Fund.
One of us (G.N.J.A.)
gratefully acknowledges the generous support from the World Laboratory.
We also thank Professor Roman Jackiw,
Professor Luis N. Epele, Professor Huner Fanchiotti,
and Professor Carlos A. Garc\'{\i}a Canal
for stimulating discussions.}

\appendix*

\section{CENTRAL POTENTIALS IN ${\bf d}$ DIMENSIONS}
\label{sec:central}

In this appendix, we summarize a few basic results for the $d$-dimensional radial
Schr\"{o}dinger equation.
We only state those properties that
are needed for the applications of the
general theory discussed in this paper.
In particular, these results are essential
for the proper use of dimensional regularization,
 even though the particular problems analyzed herein
are strictly two-dimensional.

Conservation of $d$-dimensional angular momentum
permits the separation of the radial coordinate $r$ 
from the angular variables
$
{\bf \Omega}^{(d)}  
$.
Moreover, it leads to an angular dependence
proportional to the hyperspherical
harmonics $Y_{lm} ( {\bf \Omega}^{(d)}  )$. The associated 
 wave function $\Psi ({\bf r}) =
Y_{lm}
({\bf \Omega}^{(d)})  
\,
R_{l}(r) $
includes 
a radial piece $R_{l}(r) $, which depends  upon the quantum number $l$ as well as
the energy $E$.
Furthermore,
$R_{l}(r) $
satisfies the differential equation
\begin{equation}
\left\{
\Delta_{r}^{({d})} -
\frac{l(l+ d -
2)}{r^{2}}+
\frac{2M}{\hbar^{2}}
\left[ E -V(r) \right]
\right\} R_{l}(r)
= 
0
\; ,
\label{eq:central_eff_Schr}
\end{equation}
in which the radial Laplacian
is given by 
\begin{equation}
\Delta^{({d})}_{r}
 =  \frac{1}{ r^{{d}-1} }
\frac{\partial}{\partial r}
\left( r^{{d}-1}
\frac{\partial}{\partial r} \right)
=
\frac{1}{r^{({d} -1)/2} }
\frac{\partial^{2}}{\partial r^{2}}
\left[ 
r^{({d} - 1)/2}
 \right]
-
\frac{(d- 1)(d - 3) }{4r^{2}}
\; .
\label{eq:radial_Laplacian}
\end{equation}
Then, it proves convenient to define
the reduced radial wave function
\begin{equation}
u_{l}(r)
= R_{l}(r) \,
r^{  (d -1) /2}
\;  ,
\label{eq:effective_radial_wf}
\end{equation}
which 
satisfies an effective one-dimensional Schr\"{o}dinger equation
\begin{equation}
\left\{
\frac{d^{2}}{dr^{2}}
+
\frac{2M}{\hbar^{2}}
\left[ E- V(r) 
\right]
- \frac{\Lambda_{l, {d} }}{r^{2}}
\right\}
u_{l}(r)
= 0
\;  ,
\label{eq:radial_Schr}
\end{equation}
in which
 the angular-momentum effective potential 
has a coupling constant
\begin{equation}
\Lambda_{l, {d}}
=
l(l+ d - 2)+
(d - 1)(d - 3)/4
=
 (l+\nu)^{2} - 1/4
\;  ,
\label{eq:cf_coupling}
\end{equation}
with
\begin{equation}
\nu= \frac{d}{2} -1 
\; .
\label{eq:nu_def}
\end{equation}
For the interactions discussed in this paper, Eq.~(\ref{eq:radial_Schr}) provides a direct 
transition to a differential equation of the form
\begin{equation}
\left[
\frac{d^{2}}{dr^{2}}
+
\left( 
k^{2} -
\frac{  p^{2} - 1/4 }{r^{2}}
\right)
\right]
u (r)
= 0
\;  ,
\label{eq:radial_Bessel}
\end{equation}
whose solution $u(r) = \sqrt{r} \,
Z_{p}(kr)$
is given in terms of Bessel functions  $Z_{p}(kr)$
of order $p$.

Furthermore, Eq.~(\ref{eq:radial_Schr})
should be supplemented by appropriate 
 boundary conditions 
 at the origin and at infinity.
First,
asymptotically
with respect to
$r \rightarrow \infty$,
 the bound-state solutions should have a zero limit
 in order to
satisfy the square-integrability condition.
Likewise,  
the scattering solutions are subject 
to the usual requirement
that the wave function asymptotically reproduce the incident wave plus
an outgoing scattered state~\cite{cam:dtI}.
On the other hand, the boundary condition at $r=0$
is much subtler and requires additional  study.

It turns out that the boundary condition at the origin is the key factor
that determines the nature of the singularity at the origin.
As such, it is used to establish
the classification of potentials into the regular and singular
families.
 In this framework, both the
 regular and the regularized singular interactions
are  subject to the limiting condition
\begin{equation}
r^{2} V(r)
\stackrel{ r \rightarrow 0}{\longrightarrow}
0
\;  .
\end{equation}
In particular,
asymptotically with respect to  $r \rightarrow 0$,
the wave function is reduced to the
solution of the radial part of
Laplace's equation.
As a consequence,
the regular boundary condition becomes
\begin{equation}
R_{l}(r)
\propto
r^{l}
\;  ,
\label{eq:asympt_BC}
\end{equation}
which is usually restated in terms of
 the weaker
condition
\begin{equation}
u_{ l} (0) = 0
\;  .
\label{eq:BC_at_origin}
\end{equation}

\end{document}